\documentclass[12pt]{article}
\pdfoutput=1
\usepackage{amsmath}
\usepackage{graphics,graphicx,xcolor}

\def\be{\begin{equation}}
\def\ee{\end{equation}}
\def\arr{\begin{array}{rll}}
\def\ea{\end{array}}
\def\bea{\begin{eqnarray}}
\def\eea{\end{eqnarray}}
\def\N2{$N{=}2$}

\def\>{\rangle}
\def\<{\langle}
\def\+{\dagger}
\def\={\ =\ }

\def\bal{\begin{aligned}}
\def\eal{\end{aligned}}

\begin{document}
\begin{titlepage}
\setcounter{page}{0}
\begin{center}
{\LARGE\bf  The group-theoretic approach to perfect fluid }\\
\vskip 0.5cm
{\LARGE\bf  equations with conformal symmetry }\\
\vskip 1.5cm
\textrm{\Large Anton Galajinsky \ }
\vskip 0.7cm
{\it
Laboratory of Applied Mathematics and Theoretical Physics, \\ TUSUR, Lenin ave. 40, 634050 Tomsk, Russia} \\

\vskip 0.2cm
{e-mail: a.galajinsky@tusur.ru}
\vskip 0.5cm
\end{center}

\begin{abstract} \noindent
The method of nonlinear realizations is a convenient tool for building dynamical realizations of a Lie group, which relies solely upon structure relations of the corresponding Lie algebra.
The goal of this work is to discuss advantages and limitations of the method, which is here applied to construct perfect fluid equations with conformal symmetry.
Four cases are studied in detail, which include the Schr\"odinger group, the $\ell$-conformal Galilei group, the Lifshitz group, and the relativistic conformal group.
\end{abstract}

\vspace{0.5cm}

PACS: 11.30.-j, 02.20.Sv, 47.10.A, 47.10.ab \\ \indent
Keywords: perfect fluid equations, conformal symmetry
\end{titlepage}
\renewcommand{\thefootnote}{\arabic{footnote}}
\setcounter{footnote}0

\noindent
{\bf 1. Introduction}\\

Recently, there has been a considerable effort to understand the hydrodynamic limit of the AdS/CFT-correspondence  \cite{PSS,BRSSS,BHMR}.\footnote{Literature on the subject is overwhelmingly large. For a review and further references see \cite{MR}.} The main point of the study was to establish that
the low-frequency behavior of an interacting field theory at finite temperature could be described by fluid mechanics.

The conventional formulation of fluid dynamics relies upon an expansion scheme in which the effects of viscosity and heat transfer are regarded as corrections to the perfect fluid equations (see e.g. \cite{BRSSS}).
As is known, for a properly chosen equation of state the (non)relativistic perfect fluid equations exhibit conformal invariance (see e.g. \cite{RS,JNPP,FO}). It is then natural to wonder whether such equations can be formulated on purely group-theoretic grounds.

A convenient tool for building dynamical realizations of a Lie group, which relies solely upon structure relations of the corresponding Lie algebra, is provided by the method of nonlinear realizations \cite{CWZ}.
The scheme includes several steps. First of all, one introduces coordinates and fields, whose number in general is equal to the number of generators of a Lie algebra at hand. Then one builds a formal group-theoretic element $g$, which is a product of exponentials of the type $e^{{\rm i} \alpha T}$, where $\alpha$ is a coordinate (or a field) and $T$ is a Lie algebra generator. After that, one studies the left action of the group upon the group-theoretic element $g$ and establishes transformation laws of the coordinates and fields. Finally, one computes the Maurer-Cartan one-forms $g^{-1} d g$. By construction, they hold invariant under the left action of the group upon the group theoretic element and, hence, provide convenient building blocks to formulate invariant equations of motion. If desirable, they can also be used to eliminate some of the fields from the consideration by imposing constraints. Within the method of nonlinear realizations, imposing constraints is attributed to the inverse Higgs phenomenon \cite{IO}.

An alternative formulation of fluid dynamics in terms of group-valued variables, which relies upon Kirillov's method of orbits \cite{K}, was proposed in \cite{BJLNP}.
The formalism is particularly suitable for taking into account constituent particles, which carry nonabelian charges or spin degrees of freedom, as well as for incorporating anomalies \cite{NRR}--\cite{N}.

The goal of this work is to discuss advantages and limitations of the method of nonlinear realizations, which is here applied to construct perfect fluid equations with conformal symmetry.
Four cases are studied in detail, which include the Schr\"odinger group, the $\ell$-conformal Galilei group, the Lifshitz group, and the relativistic conformal group.

The organization of the work is as follows. In the next section, we briefly outline key features of the method of nonlinear realizations, which are used in later section to explore fluid equations with conformal symmetry.
For simplicity of the presentation, the construction is illustrated by the example of $so(2,1)$ algebra, which was first studied in \cite{IKL}. In Sect. 3, perfect fluid equations with the Schr\"odinger symmetry are built in terms of the invariant Maurer-Cartan one-forms associated with the Schr\"odinger group and an invariant derivative. It is demonstrated that a proper equation of states, which links pressure to fluid density, comes about quite naturally without the need to invoke more sophisticated arguments \cite{RS,JNPP}. In Sect. 4 and Sect. 5, a similar analysis is performed for the $\ell$-conformal Galilei group and the Lifshitz group, respectively.
To the best of our knowledge, the Lifshitz-invariant equations in Sect. 5 are new.  Sect. 6 is focused on the case of the relativistic conformal group. In contrast to the nonrelativistic examples, the construction of relativistic fluid equations invariant under the conformal group in terms of the Maurer-Cartan invariants alone turns to be problematic and extra arguments need to be invoked. In the concluding Sect. 7, we summarize our results and discuss possible further developments.

Throughout the paper, summation over repeated indices is understood unless otherwise is stated.

\vspace{0.5cm}

\noindent
{\bf 2. The method of nonlinear realizations}\\

In this section, we outline key features of the method of nonlinear realizations \cite{CWZ}, which will be used below to explore fluid equations with conformal symmetry. For simplicity of the presentation, the construction will be illustrated by the example of $so(2,1)$ algebra
\be\label{CA}
[H,D]={\rm i} H, \qquad  [H,K]=2 {\rm i} D, \qquad [D,K]={\rm i} K,
\ee
which was first studied in \cite{IKL}. Above $H$ is interpreted as the temporal translation generator, $D$ links to dilatation, and $K$ is associated with the special conformal transformation.

In its essence, the method of nonlinear realizations is a tool to build dynamical realizations of a Lie group, which is based solely upon structure relations of the corresponding Lie algebra. As the first step of the construction, one introduces coordinates and fields whose number in general is equal to the number of generators of the Lie algebra. The choice is not unique and it essentially depends on a dynamical realization one seeks for.  For example, if one is concerned with one-dimensional mechanics originating from (\ref{CA}), a temporal variable $t$ and a function $u(t)$, which describes a particle dynamics, are needed. It seems natural to link $t$ to $H$ and $u(t)$ to $D$. In order to treat all the generators on equal footing, one introduces one more field $w(t)$, which is regarded as a partner of $K$. At this stage, it is not yet clear whether $u(t)$ or $w(t)$ will be more suitable for describing a reasonable conformal mechanics model and one anticipates an $SO(2,1)$-invariant constraint, which will link the fields to each other.

Then one introduces the group-theoretic element
\be\label{ECA}
g=e^{{\rm i} t H} e^{{\rm i} u(t) D} e^{{\rm i} w(t) K} .
\ee
In general, the expressions in (\ref{CA}) are regarded as formal Lie brackets (rather than commutators of operators) and the exponential entering (\ref{ECA}) is treated as the exponential map of a Lie algebra to a neighborhood of the unit group element. Yet, nothing prevents one from assuming that a specific representation of the Lie algebra is chosen such that (\ref{CA}) represents commutators of the generators and (\ref{ECA}) is an operator acting upon a state. In the latter case, the exponential entering (\ref{ECA}) can be regarded as a formal Taylor series $e^{A}=\sum_{n=0}^\infty\frac{A^n}{n!} $, in which case the well known Baker-Campbell-Hausdorff formula
\be\label{ser}
e^{iA}~ T~ e^{-iA}=T+\sum_{n=1}^\infty\frac{i^n}{n!}
\underbrace{[A,[A, \dots [A,T] \dots]]}_{n~\rm times}
\ee
is applicable, where $A$ and $T$ are two arbitrary operators. The conventional property of the exponential
\be
e^{\alpha A} e^{\beta A}=e^{(\alpha+\beta) A},
\ee
holds true as well, where $\alpha$, $\beta$ are constants and $A$ is an operator action upon a state in the representation chosen. Eq. (\ref{ser}) is the cornerstone of the whole consideration to follow. Note that the order in which the factors contribute to the group-theoretic element (\ref{ECA}) can be chosen at will and different options are related by coordinate and field redefinitions.

At the second step of the construction, one analyzes the left action of the group upon the group-theoretic element
\be\label{STR}
g'=e^{{\rm i} \beta H} e^{{\rm i} \lambda D} e^{{\rm i} \sigma  K}  \cdot g=e^{{\rm i} t' H} e^{{\rm i} u'(t') D} e^{{\rm i} w'(t') K},
\ee
where $\beta$, $\lambda$, $\sigma$ are real transformation parameters, and makes use of (\ref{ser}) so as to establish transformation laws of the coordinates and fields. For most physical applications it suffices to consider their infinitesimal form.

When performing calculations, depending on a specific operator $e^{L}$ at hand, it might prove helpful to insert the unit operator $e^{-B} e^{B}$, with $B$ to be specified below, either to the left or to the right of $e^{L}$. The following chain of relations
\bea\label{SL}
&&
e^{{\rm i} \lambda D}  e^{{\rm i} t H} =e^{{\rm i} \lambda D}  e^{{\rm i} t H}  \underbrace{ e^{-{\rm i} \lambda D}  e^{{\rm i} \lambda D}}_{\bf{1}}=e^{{\rm i} \lambda D} \left(1+{\rm i} t H+\dots \right) e^{-{\rm i} \lambda D}  e^{{\rm i} \lambda D}=
\nonumber\\[2pt]
&&
 \left(1+{\rm i} t \left(H+[{\rm i} \lambda D,H]+\dots \right)   +\dots \right) e^{{\rm i} \lambda D}=e^{{\rm i} (1+\lambda)t H} e^{{\rm i} \lambda D}
\eea
gives the example of how $e^{{\rm i} \lambda D}$ "passes through" $e^{{\rm i} t H}$. The Baker-Campbell-Hausdorff formula was repeatedly applied in the second line of (\ref{SL}). At the same time, when computing $e^{{\rm i} \sigma  K} e^{{\rm i} t H}$ with infinitesimal $\sigma$, it proves convenient to place $e^{{\rm i} t H} e^{-{\rm i} t H}$ to the left of the operator
\bea
&&
e^{{\rm i} \sigma  K} e^{{\rm i} t H}=\underbrace{ e^{{\rm i} t H} e^{-{\rm i} t H} }_{\bf{1}} e^{{\rm i} \sigma  K} e^{{\rm i} t H}=e^{{\rm i} t H} e^{-{\rm i} t H} \left(1+ {\rm i} \sigma  K \right)e^{{\rm i} t H}=
\nonumber\\[2pt]
&&
e^{{\rm i} t H} \left(1+{\rm i} \sigma \left(K-[{\rm i} t H,K]+\frac{1}{2!} [{\rm i} t H,[{\rm i} t H,K]] \right) \right)=e^{{\rm i}(t+\sigma t^2)H} e^{{\rm i} \sigma K} e^{2{\rm i} \sigma t D},
\eea
where we used the fact that $e^{{\rm i} \sigma A} e^{{\rm i} \sigma B}=e^{{\rm i} \sigma B} e^{{\rm i} \sigma A}$ for infinitesimal $\sigma$.

Taking into account the technicalities, from Eq. (\ref{STR}) one obtains the infinitesimal $SO(2,1)$-transformations acting upon the temporal variable and the fields $u(t)$, $w(t)$ (each transformation is separated by semicolon)
\begin{align}\label{STR1}
&
t'=t+\beta, && u'(t')=u(t), && w'(t')=w(t);
\nonumber\\[6pt]
&
t'=(1+\lambda)t, && u'(t')=u(t)+\lambda, && w'(t')=w(t);
\nonumber\\[2pt]
&
t'=t+\sigma t^2, && u'(t')=u(t)+2\sigma t, && w'(t')=w(t)+e^{u(t)} \sigma.
\end{align}

The third step of the method consists in computing the Maurer-Cartan one-forms\footnote{Given $g$ in (\ref{ECA}), the inverse element is $g^{-1}=e^{-{\rm i} w K} e^{-{\rm i} u D} e^{-{\rm i} t H}$, while the differential reads $dg=e^{{\rm i} t H} \left( {\rm i} dt H \right)e^{{\rm i} u D} e^{{\rm i} w K}+e^{{\rm i} t H} e^{{\rm i} u D} \left( {\rm i} du D \right) e^{{\rm i} w K}+e^{{\rm i} t H} e^{{\rm i} u D} e^{{\rm i} w K} \left( {\rm i} dw K \right)$.}
\be
g^{-1} d g={\rm i} \omega_H H+ {\rm i} \omega_D D+{\rm i} \omega_K K,
\ee
where
\be\label{SMCI}
\omega_H=e^{-u} dt, \qquad \omega_D=du-2 w e^{-u} dt, \qquad \omega_K=dw-w du+w^2 e^{-u} dt.
\ee
By construction, they hold invariant under the group transformation (\ref{STR}), (\ref{STR1}) and, hence, provide convenient building blocks to formulate invariant equations of motion. If desirable, they can also be used to eliminate some of the fields entering the group-theoretic element from the consideration by imposing constraints.

In the last step of the construction, one specifies a dynamical realization of the group at hand by choosing judiciously a combination of the Maurer-Cartan invariants which results in a reasonable set of second order differential equations. The latter are identified with the equations of motion of a dynamical system. This item is not straightforward and requires guesswork. For example, if one wishes to use (\ref{SMCI}) in order to formulate one-dimensional $SO(2,1)$-invariant mechanics, one imposes the constraint $\omega_D=0$, which links $w$ to $u$
\be
w=\frac 12 \frac{d e^u}{dt} ,
\ee
and then postulates the equation of motion  \cite{IKL}
\be\label{DU}
\omega_K-\gamma^2 \omega_H=0,
\ee
in which $\gamma$ is a (coupling) constant. Implementing the field redefinition $u=\ln \rho^2$, which is meant to remove the ${\left(\frac{d u}{dt} \right)}^2$-term from (\ref{DU}), one finally arrives at
\be\label{DFF}
\frac{d^2 \rho}{dt^2}=\frac{\gamma^2}{\rho^3},
\ee
which is the conventional $1d$ conformal mechanics equation of motion \cite{DFF}.

To give another example, it is known that the Schwarzian derivative
\be\label{SD}
S(\rho(t))=\frac{\dddot{\rho}(t)}{\dot\rho(t)}-\frac 32 {\left(\frac{\ddot{\rho}(t)}{\dot\rho(t)}\right)}^2,
\ee
where $\rho(t)$ is a real function and the dot designates the derivative with respect to $t$, holds invariant under the $SL(2,R)$-transformation acting upon the argument
\be
\rho'(t)=\frac{a \rho(t)+b}{c \rho(t)+d},
\ee
with $ad-cb=1$. At first glance, the invariance does not seem obvious at all. Yet, because $sl(2,R)\sim so(2,1)$, one can naturally arrive at (\ref{SD}) by applying the group-theoretic arguments similar to those above \cite{AG1}. It suffices to keep the temporal variable $t$ as an external parameter\footnote{To be more precise, one considers $R \times SL(2,R)$ and links $t$ to the generator of the group of real numbers under addition $R$.} and introduce the group-theoretic element
\be
g=e^{{\rm i} \rho(t)H} e^{{\rm i} s(t) K} e^{{\rm i} u(t) D},
\ee
where $\rho(t)$, $s(t)$, $u(t)$ are as yet unspecified functions, which gives rise to the Maurer-Cartan invariants
\be\label{MCI}
\omega_H=\dot\rho e^{-u} dt, \qquad \omega_K=e^u \left(\dot s+s^2 \dot\rho\right)dt, \qquad \omega_D=\left(\dot u-2 s \dot\rho\right) dt.
\ee
Imposing the constraints $\omega_H-\mu dt=0$ and $\omega_D+2 \nu dt=0$, where $\mu$ and $\nu$ are arbitrary constants, one can express $u$ and $s$ in terms of $\rho$, while substituting the result into the only remaining one-form $\omega_K$, one arrives at the Schwarzian derivative \cite{AG1}.

Our primary concern in later sections will be to understand whether the construction outlined above is powerful enough to result in perfect fluid equations with conformal symmetry. Specifically, the cases of the Schr\"odinger group, the $\ell$-conformal Galilei group, the Lifshitz group, and the relativistic conformal group will discussed in turn.

\vspace{0.5cm}

\noindent
{\bf 3. Perfect fluid equations with the Schr\"odinger symmetry}\\

The Lie algebra associated with the Schr\"odinger group includes generators of temporal translation, dilatation, and special
conformal transformation, which form $so(2,1)$ subalgebra, as well as spatial translations, the Galilei boosts, and spatial rotations. Given a nonrelativistic spacetime parameterized by a temporal variable $t$ and Cartesian coordinates $x_i$, $i=1,\dots,\mathcal{N}$, where $\mathcal{N}$ is the spatial dimension, they can be represented in the form\footnote{Here and in what follows the conventional rotation generators are disregarded.}
\begin{align}
&
H={\rm i} \frac{\partial}{\partial t}, &&  D={\rm i}  \left(t \frac{\partial}{\partial t}+\frac 12 x_i \frac{\partial}{\partial x_i} \right), && K={\rm i}  \left(t^2 \frac{\partial}{\partial t}+ t x_i \frac{\partial}{\partial x_i}\right),
\nonumber\\[2pt]
&
C^{(0)}_i={\rm i} \frac{\partial}{\partial x_i}, && \qquad C^{(1)}_i={\rm i} t \frac{\partial}{\partial x_i}, &&
\end{align}
which obey the commutation relations
\begin{align}\label{SA}
&
[H,D]={\rm i} H, &&  [H,K]=2 {\rm i} D, && [D,K]={\rm i} K, && [H,C^{(1)}_i]={\rm i} C^{(0)}_i,
\nonumber\\[2pt]
&
[D,C^{(0)}_i]=-\frac{{\rm i}}{2} C^{(0)}_i, && [D,C^{(1)}_i]=\frac{{\rm i}}{2} C^{(1)}_i, && [K,C^{(0)}_i]=-{\rm i} C^{(1)}_i. &&
\end{align}
The finite form of the corresponding transformations acting in the nonrelativistic spacetime reads
\bea\label{Sch}
&&
t'=\frac{\alpha t+\beta}{\gamma t+\delta}, \qquad
x'_i={\left(\frac{\partial t'}{\partial t} \right)}^{\frac 12} x_i; \qquad t'=t, \qquad x'_i=x_i+a^{(0)}_i+a^{(1)}_i t,
\eea
where $\alpha$, $\beta$, $\gamma$, $\delta$, $a^{(0)}_i$, $a^{(1)}_i$ are transformation parameters, the first four of which obey the restriction $\alpha \delta-\beta \gamma=1$.

In order to describe a fluid, one introduces the density\footnote{Throughout the paper, we use units in which $\rho$ is dimensionless. This can be done by choosing a reference value $\rho_0$ and rescaling $\rho \to \frac{\rho}{\rho_0}$.} $\rho(t,x)$ and the velocity vector field $\upsilon_i (t,x)$, $i=1,\dots,\mathcal{N}$. The transformation law of $\rho(t,x)$ under the Schr\"odinger group is obtained by fixing a value of the temporal variable $t$ and demanding the mass of an $\mathcal{N}$-dimensional volume element $V$ to be invariant under (\ref{Sch})
\be\label{TRLR}
\int_{V'} d x' \rho' (t',x')=\int_{V} d x \rho(t,x),
\ee
where $dx=dx_1 \dots dx_\mathcal{N}$.
This gives
\be\label{trR}
\rho(t,x)= {\left(\frac{\partial t'}{\partial t} \right)}^{\frac{\mathcal{N}}{2}} \rho' (t',x'); \qquad \rho(t,x)=\rho' (t',x'),
\ee
where the first relation corresponds to the $SO(2,1)$-transformations, while the second links to the spatial translations and the Galilei boosts.

Considering an orbit of a liquid particle parameterized by $x_i (t)$, for which
\be\label{vel}
\frac{d x_i (t)}{d t}= \upsilon_i (t,x(t)),
\ee
from Eq. (\ref{Sch}) one can determine the transformation laws of the velocity vector field\footnote{As is usual in classical dynamics, when restricting transformations similar to (\ref{Sch}) to a particle orbit, one replaces $x'_i$ with $x'_i (t')$ and $x_i$ with $x_i (t)$.}
\be\label{trV}
\upsilon_i (t,x)={\left(\frac{\partial t'}{\partial t} \right)}^{\frac 12} \upsilon'_i (t',x')+\frac{\partial}{\partial t} {\left(\frac{\partial t'}{\partial t} \right)}^{-\frac 12} x'_i; \qquad \upsilon_i (t,x)=\upsilon'_i (t',x')-a^{(1)}_i,
\ee
where again the former equality corresponds to the $SO(2,1)$-transformations, while the latter links to the spatial translations and the Galilei boosts.

In what follows, we will need infinitesimal form of the Schr\"odinger transformations acting upon the coordinates and fields. Substituting $\alpha=1$, $\delta=1$, $\gamma=0$  into (\ref{Sch}), (\ref{trR}), (\ref{trV}) and regarding $\beta$ as infinitesimal parameter, one obtains the infinitesimal form of the temporal translation
\begin{align}\label{TR1}
&
t'=t+\beta, && x'_i=x_i,
\nonumber\\[2pt]
&
\rho'(t',x')=\rho(t,x), && \upsilon'_i(t',x')=\upsilon_i (t,x).
\end{align}

Choosing $\alpha=e^{\frac{\lambda}{2}}$, $\delta=e^{-\frac{\lambda}{2}}$, $\beta=0$, $\gamma=0$, setting $\lambda$ to be infinitesimal parameter and Taylor expanding in $\lambda$ up to the first order, one gets the dilatation transformation
\begin{align}\label{TR2}
&
t'=(1+\lambda) t, && x'_i=\left(1+\frac{\lambda}{2}\right) x_i,
\nonumber\\[2pt]
&
\rho'(t',x')=\left(1-\frac{ \mathcal{N} \lambda}{2}\right) \rho(t,x), && \upsilon'_i(t',x')=\left(1-\frac{\lambda}{2}\right)\upsilon_i (t,x),
\end{align}
where, as above, $\mathcal{N}$ denotes the spatial dimension.

Infinitesimal form of the special conformal transformation is found by setting $\alpha=1$, $\delta=1$, $\beta=0$, $\gamma=-\sigma$ in Eq. (\ref{Sch}), regarding $\sigma$ as infinitesimal parameter, and Taylor expanding in $\sigma$ up to the first order
\begin{align}\label{TR3}
&
t'=t+\sigma t^2, && x'_i=(1+\sigma t)  x_i,
\nonumber\\[2pt]
&
\rho'(t',x')=(1-\mathcal{N} \sigma t) \rho(t,x), && \upsilon'_i(t',x')=(1-\sigma t)\upsilon_i (t,x)+\sigma  x_i.
\end{align}

The spatial translations and the Galilei boosts maintain their form
\begin{align}\label{TR4}
&
t'=t, && x'_i=x_i+a^{(0)}_i +a^{(1)}_i t,
\nonumber\\[2pt]
&
\rho'(t',x')=\rho(t,x), && \upsilon'_i(t',x')=\upsilon_i (t,x)+a^{(1)}_i.
\end{align}

Having fixed transformation laws of the fields which characterize a fluid,
we are now in a position to construct invariant equations of motion within the method of nonlinear realizations.
As the first step, one has to introduce a proper group-theoretic element.
The building blocks, which are at our disposal, are the coordinates $t$, $x_i$ and the fields $\rho(t,x)$, $\upsilon_i (t,x)$ on the one hand, and the generators of the Schr\"odinger algebra $H$, $D$, $K$, $C^{(0)}_i$, $C^{(1)}_i$ on the other hand. It seems natural to link the generator of temporal translation $H$ and the generator of spatial translations $C^{(0)}_i$ to the temporal variable $t$ and the Cartesian coordinates $x_i$, respectively.
The remaining generators of the algebra should be accompanied by specific combinations of $\rho(t,x)$ and $\upsilon_i (t,x)$ in a way compatible with the left action of the Schr\"odinger group upon a group-theoretic element, which will be introduced shortly. Because $C^{(1)}_i$ carries a vector index, its partner in a group-theoretic element should be proportional to the velocity vector field $\upsilon_i (t,x)$. For symmetry reasons (see below), a coefficient of proportionality can be taken to be just the unity.  Denoting companions of $D$, $K$ by $u(t,x)$, $w(t,x)$, respectively, and postponing for later a clarification of their connection to $\rho(t,x)$, $\upsilon_i (t,x)$, one finally gets the group-theoretic element
\be\label{gte}
g=e^{{\rm i} t H} e^{{\rm i} x_i C^{(0)}_i} e^{{\rm i}  \upsilon_i (t,x) C^{(1)}_i} e^{{\rm i} u(t,x) D} e^{{\rm i} w(t,x) K}.
\ee
As was mentioned is Sect. 2, factors contributing to a group-theoretic element in general do not commute. The choice (\ref{gte}) proves to be most convenient for our subsequent analysis.

As the next step, one considers the left action of the Schr\"odinger group upon the group-theoretic element
\bea
&&
e^{{\rm i} \beta H} e^{{\rm i} a^{(0)}_i C^{(0)}_i} e^{{\rm i}  a^{(1)}_i C^{(1)}_i} e^{{\rm i} \lambda D} e^{{\rm i} \sigma K} \cdot e^{{\rm i} t H} e^{{\rm i} x_i C^{(0)}_i} e^{{\rm i}  \upsilon_i (t,x) C^{(1)}_i} e^{{\rm i} u(t,x) D} e^{{\rm i} w(t,x) K}=
\nonumber\\[2pt]
&&
e^{{\rm i} t' H} e^{{\rm i} x'_i C^{(0)}_i} e^{{\rm i}  \upsilon'_i (t',x') C^{(1)}_i} e^{{\rm i} u'(t',x') D} e^{{\rm i} w'(t',x') K},
\eea
where $\beta$, $\lambda$, $\sigma$, $a^{(0)}_i$, $a^{(1)}_i$ are transformation parameters,
and makes use of the Baker-Campbell-Hausdorff formula (\ref{ser}) to determine transformation laws of the coordinates and fields. For simplicity of the presentation, we focus on their infinitesimal form (each transformation is separated by semicolon)
\begin{align}\label{TR5}
&
t'=t+\beta, && x'_i=x_i, &&
\nonumber\\[2pt]
&
\upsilon'_i (t',x')=\upsilon_i (t,x), && u'(t',x')=u(t,x), && w'(t',x')=w(t,x);
\nonumber\\[6pt]
&
t'=(1+\lambda)t, && x'_i=\left(1+\frac{\lambda}{2}\right)x_i, &&
\nonumber\\[2pt]
&
\upsilon'_i (t',x')=\left(1-\frac{\lambda}{2}\right) \upsilon_i (t,x), && u'(t',x')=u(t,x)+\lambda, && w'(t',x')=w(t,x);
\nonumber\\[6pt]
&
t'=t+\sigma t^2, && x'_i=(1+\sigma t) x_i, &&
\nonumber\\[2pt]
&
\upsilon'_i (t',x')=(1-\sigma t)\upsilon_i (t,x)+\sigma x_i, && u'(t',x')=u(t,x)+2\sigma t, && w'(t',x')=w(t,x)+\sigma e^{u(t,x)};
\nonumber\\[6pt]
&
t'=t, && x'_i=x_i+a^{(0)}_i, &&
\nonumber\\[2pt]
&
\upsilon'_i (t',x')=\upsilon_i (t,x), && u'(t',x')=u(t,x), && w'(t',x')=w(t,x);
\nonumber\\[6pt]
&
t'=t, && x'_i=x_i+t a^{(1)}_i, &&
\nonumber\\[2pt]
&
\upsilon'_i (t',x')=\upsilon_i (t,x)+a^{(1)}_i, && u'(t',x')=u(t,x), && w'(t',x')=w(t,x).
\end{align}

At this stage, comparing (\ref{TR5}) with (\ref{TR1}), (\ref{TR2}), (\ref{TR3}), (\ref{TR4}) and taking into account the identities
\be\label{Id}
\frac{\partial}{\partial t}=\left(\frac{\partial t'}{\partial t}\right) \frac{\partial}{\partial t'}+\left(\frac{\partial x'_i}{\partial t} \right) \frac{\partial}{\partial x'_i},
\qquad \frac{\partial}{\partial x_i}=\left( \frac{\partial t'}{\partial x_i} \right) \frac{\partial}{\partial t'}+\left(\frac{\partial x'_j}{\partial x_i} \right)\frac{\partial}{\partial x'_j},
\ee
one can unambiguously link $u$, $w$ to $\rho$ and $\upsilon_i$
\be\label{rw}
\rho=e^{-\frac{\mathcal{N}}{2} u}, \qquad w=\frac{1}{\mathcal{N}} \rho^{-\frac{2}{\mathcal{N}}} \frac{\partial \upsilon_i }{\partial x_i}.
\ee
A comparison of (\ref{TR5}) with (\ref{TR1}), (\ref{TR2}), (\ref{TR3}), (\ref{TR4}) also confirms our earlier identification of a companion of the Galilei boost generator $C^{(1)}_i$ entering the group-theoretic element with the fluid velocity vector field $\upsilon_i$.

As the final step, one computes the invariant Maurer-Cartan one-forms
\be
g^{-1} d g={\rm i} \omega_H  H +{\rm i} \omega_D D +{\rm i} \omega_K K +{\rm i} \omega^{(0)}_i C^{(0)}_i  +{\rm i} \omega^{(1)}_i C^{(1)}_i,
\ee
where
\bea\label{MC}
&&
\omega_H=e^{-u} dt, \qquad \omega_D=du-2 w e^{-u} dt, \qquad \omega_K=dw-w du+ w^2 e^{-u} dt,
\nonumber\\[2pt]
&&
\omega^{(0)}_i=e^{-\frac{u}{2}}(dx_i-\upsilon_i dt), \qquad \omega^{(1)}_i=e^{\frac{u}{2}} d \upsilon_i- w e^{-\frac{u}{2}} (dx_i-\upsilon_i dt),
\eea
and takes notice of the fact that the derivative
\be\label{Na}
\nabla_i=\rho^{-\frac{1}{\mathcal{N}}} \frac{\partial }{\partial x_i},
\ee
is invariant under the action of the Schr\"odinger group.\footnote{The invariance of the derivative is most easily verified by making recourse to the finite transformations (\ref{Sch}), (\ref{trR}) and taking into account the identity (\ref{Id}).}

The invariants (\ref{MC}) and (\ref{Na}) are all one needs to formulate perfect fluid equations with the Schr\"odinger symmetry within the group-theoretic approach. First of all, a comparison of $\omega^{(0)}_i$ with one of the key ingredients of fluid mechanics (\ref{vel}) suggests a further specification in the group-theoretic element (\ref{gte})
\be\label{xtox}
x_i \to x_i(t),
\ee
where $x_i(t)$ is now interpreted as parameterizing an orbit of a liquid particle. The substitution (\ref{xtox}) also links the differential $d$ to the material derivative $\mathcal{D}$, which is commonly used within fluid mechanics
\be\label{MD}
d=dt \mathcal{D}, \qquad \mathcal{D}=\frac{\partial}{\partial t} +\upsilon_i  (t,x) \frac{\partial}{\partial x_i}.
\ee
Imposing the Schr\"odinger-invariant constraint
\be\label{c1}
\omega^{(0)}_i=0,
\ee
one thus reproduces Eq. (\ref{vel}) and confirms once again that the identification of $\upsilon_i$ in (\ref{gte}) with the fluid velocity vector field was correct.

Substituting (\ref{rw}) into $\omega_D$, one gets
\be
\omega_D=-\frac{2 }{\mathcal{N}} \left(\frac{\mathcal{D \rho}}{\rho} +\frac{\partial \upsilon_i }{\partial x_i}\right)  dt.
\ee
Hence, demanding $\omega_D$ to vanish, one naturally arrives at the continuity equation
\be\label{CE}
\frac{\partial \rho}{\partial t} + \frac{\partial ( \rho \upsilon_i )}{\partial x_i}=0,
\ee
which is the equation of motion for $\rho$ in fluid mechanics.

It remains to determine the equation for $\upsilon_i$. Note that after imposing the constraint (\ref{c1}), the one-form $\omega^{(1)}_i$ simplifies to
\be
\omega^{(1)}_i=\left( \rho^{-\frac{1}{\mathcal{N}}} \mathcal{D} \upsilon_i \right) dt.
\ee
In particular, it involves the material derivative of the velocity vector field $\upsilon_i$,  which is the fluid mechanics analog of the acceleration vector in Newtonian mechanics. Because the remaining Maurer-Cartan forms in (\ref{MC}) do not carry vector indices, one is led to use the invariant derivative (\ref{Na}) so as to specify a potential term. Imposing the simplest Schr\"odinger-invariant restriction
\be
\omega^{(1)}_i+\alpha \nabla_i \omega_H=0,
\ee
where $\alpha$ is a real constant, one gets the equation of motion
\be
\mathcal{D} \upsilon_i =-\alpha \frac{\partial \rho^{\frac{2}{\mathcal{N}}}}{\partial x_i}.
\ee
The latter can be put into the conventional Euler form
\be \label{EE}
\rho \mathcal{D} \upsilon_i =-\frac{\partial p}{\partial x_i},
\ee
by introducing the pressure $p(t,x)$ which obeys the equation of state
\be\label{ES}
p=\nu \rho^{1+\frac{2}{\mathcal{N}}},
\ee
with $\nu=\frac{2\alpha}{\mathcal{N}+2}$. Eqs. (\ref{CE}), (\ref{EE}), (\ref{ES}) reproduce the perfect fluid equations invariant under the action of the Schr\"odinger group, which were originally introduced in \cite{RS,JNPP} (for conserved charges and the energy-momentum tensor see \cite{JNPP}). Within the group-theoretic approach, they result from imposing the invariant constraints $\omega^{(0)}_i=0$, $\omega_D=0$, $\omega^{(1)}_i+\alpha \nabla_i \omega_H=0$
upon the Maurer-Cartan one-forms. The advantage of the method is that the equation of state (\ref{ES}) comes about quite naturally without the need to invoke more sophisticated arguments (cf. \cite{RS,JNPP}).

Concluding this section, we note that some statements in the literature regarding conformal symmetries of nonrelativistic fluid mechanics contradict each other. For a detailed account and further references see \cite{HZ}.

\vspace{0.5cm}

\noindent
{\bf 4. Perfect fluid equations with the $\ell$-conformal Galilei symmetry}\\

As is well known, the Schr\"odinger algebra is a particular instance of the so called $\ell$--conformal Galilei algebra \cite{Henkel,NOR}. The latter includes generators of
(temporal) translation, dilatation, special
conformal transformation, spatial rotations, spatial translations, Galilei boosts and constant accelerations. They obey the structure relations\footnote{As above, in this section we disregard spatial rotations.}
\begin{align}\label{algebra}
&
[H,D]={\rm i} H, &&  [H,K]=2 {\rm i}  D, && [D,K]={\rm i}  K,
\nonumber\\[2pt]
&
[H,C^{(n)}_i]={\rm i}  n C^{(n-1)}_i, && [D,C^{(n)}_i]={\rm i}  (n-l) C^{(n)}_i, && [K,C^{(n)}_i]={\rm i}  (n-2l) C^{(n+1)}_i,
\end{align}
where $i=1,\dots,\mathcal{N}$ and $n=0,\dots, 2\ell$. Here $\mathcal{N}$ designates the spatial dimension and $\ell$ is an arbitrary (half)integer real parameter, which gives the name to the algebra.

A conventional realization of the algebra in terms of differential operators acting in a nonrelativistic spacetime reads
\be
H={\rm i} \frac{\partial}{\partial t}, \quad D={\rm i} \left(t \frac{\partial}{\partial t}+\ell x_i \frac{\partial}{\partial x_i} \right), \quad K={\rm i} \left(t^2 \frac{\partial}{\partial t}+2 \ell t x_i \frac{\partial}{\partial x_i} \right), \quad C^{(n)}_i={\rm i} t^n  \frac{\partial}{\partial x_i}.
\ee
Note that the value of the parameter $\ell$ specifies the number of acceleration generators at hand ($C^{(n)}_i$ with $n>1$) and for $\ell=\frac 12$ one reveals the Schr\"odinger algebra discussed in the preceding section.
A finite form of the transformations is given by (no sum over repeated index $n$)
\be\label{TTR1}
t'=\frac{\alpha t+\beta}{\gamma t+\delta}, \qquad
x'_i={\left(\frac{\partial t'}{\partial t} \right)}^\ell x_i; \qquad t'=t, \qquad x'_i=x_i+a^{(n)}_i t^n,
\ee
where $n=0,\dots,2\ell$, and $\alpha$, $\beta$, $\gamma$, $\delta$, $a^{(0)}_i$, $a^{(1)}_i$, $\dots$, $a^{(2\ell)}_i$ are transformation parameters, the first four of which obey the restriction $\alpha \delta-\beta \gamma=1$.

In a very recent work \cite{AG2}, the perfect fluid equations with the Schr\"odinger symmetry were generalized so as to accommodate the $\ell$--conformal Galilei symmetry
\be\label{fin}
\frac{\partial \rho}{\partial t} + \frac{\partial ( \rho \upsilon_i )}{\partial x_i}=0, \qquad \rho  \mathcal{D}^{2\ell} \upsilon_i=-\frac{\partial p}{\partial x_i}, \qquad p=\nu \rho^{1+\frac{1}{\ell \mathcal{N}}},
\ee
where $\nu$ is a constant. Conserved charges and the energy-momentum tensor were built as well \cite{AG2}. Our goal in this section, is to demonstrate that the group-theoretic approach allows one to arrive at (\ref{fin}) in a rather natural and efficient way.

As the first step, one has to determine transformation laws of the fluid density and the velocity vector field under the action of the $\ell$-conformal Galilei group. Repeating the arguments in the preceding section, one finds (no sum over repeated index $n$)
\begin{align}\label{TTR2}
&
\rho(t,x)= {\left(\frac{\partial t'}{\partial t} \right)}^{\ell \mathcal{N}} \rho' (t',x'); && \upsilon_i (t,x)={\left(\frac{\partial t'}{\partial t} \right)}^{1-\ell} \upsilon'_i (t',x')+\frac{\partial}{\partial t} {\left(\frac{\partial t'}{\partial t} \right)}^{-\ell} x'_i;
\nonumber\\[2pt]
&
\rho(t,x)=\rho' (t',x'), && \upsilon_i (t,x)=\upsilon'_i (t',x')-n a^{(n)}_i t^{n-1},
\end{align}
where the first line corresponds to the $SO(2,1)$-transformations, while the second line describes the accelerations. Note that, similarly to the Schr\"odinger case, the derivative
\be\label{Na1}
\nabla_i=\rho^{-\frac{1}{\mathcal{N}}} \frac{\partial }{\partial x_i},
\ee
holds invariant under the transformations (\ref{TTR1}) and (\ref{TTR2}).
Eqs. (\ref{TTR2}) will be used below in order to link fields contributing to a group-theoretic element to $\rho$ and $\upsilon_i$, while (\ref{Na1}) will help to formulate invariant equations of motion.

Then one considers a natural generalization of the group theoretic element (\ref{gte})
\be\label{GTE}
g=e^{{\rm i} t H} e^{{\rm i} x_i C^{(0)}_i} e^{{\rm i}  \upsilon^{(1)}_i (t,x) C^{(1)}_i}  \dots e^{{\rm i}  \upsilon^{(2\ell)}_i (t,x) C^{(2\ell)}_i} e^{{\rm i} u(t,x) D} e^{{\rm i} w(t,x) K},
\ee
where $t$ and $x_i$ are the temporal and spatial coordinates, $\upsilon^{(1)}_i$ is identified with the fluid velocity vector field $\upsilon_i$, new vector fields $\upsilon^{(2)}_i$, $\dots$, $\upsilon^{(2\ell)}_i$ will be later linked to the material derivatives of $\upsilon_i$, whereas $u$, $w$ are scalars to be expressed in terms of $\rho$ and $\upsilon_i$ in accord with the way in which they transform under the $\ell$-conformal Galilei group.

Analyzing the left action of the group  upon the element (\ref{GTE})
\bea
&&
g'=e^{{\rm i} \beta H} e^{{\rm i} a^{(0)}_i C^{(0)}_i} e^{{\rm i}  a^{(1)}_i (t,x) C^{(1)}_i} \dots e^{{\rm i}  a^{(2\ell)}_i (t,x) C^{(2\ell)}_i} e^{{\rm i} \lambda D} e^{{\rm i} \sigma  K} \cdot g,
\eea
where $\beta$, $\lambda$, $\sigma$, $a^{(0)}_i$, $\dots$, $a^{(2\ell)}_i$ are infinitesimal transformation parameters, one gets (each transformation is separated by semicolon; no sum over repeated indices $k$ and $s$)
\begin{align}\label{TR6}
&
t'=t+\beta, && x'_i=x_i, &&
\nonumber\\[2pt]
&
\upsilon'^{(k)}_i (t',x')=\upsilon_i^{(k)} (t,x), && u'(t',x')=u(t,x), && w'(t',x')=w(t,x);
\nonumber\\[6pt]
&
t'=(1+\lambda)t, && x'_i=\left(1+\lambda \ell \right)x_i, &&
\nonumber\\[2pt]
&
\upsilon'^{(k)}_i (t',x')=\left(1-\lambda(k-\ell)\right) \upsilon^{(k)}_i (t,x), && u'(t',x')=u(t,x)+\lambda, && w'(t',x')=w(t,x);
\nonumber\\[6pt]
&
t'=t+\sigma t^2, && x'_i=(1+2 \sigma \ell t) x_i, &&
\nonumber\\[2pt]
&
\upsilon'^{(k)}_i (t',x')=(1-2 \sigma (k-\ell) t)\upsilon^{(k)}_i (t,x) && u'(t',x')=u(t,x) && w'(t',x')=w(t,x)
\nonumber\\[2pt]
&
\qquad \qquad ~  -\sigma (k-1-2\ell) \upsilon^{(k-1)}_i (t,x), && \qquad \qquad ~ +2\sigma t && \qquad \quad ~+\sigma e^{u(t,x)};
\nonumber\\[6pt]
&
t'=t, && x'_i=x_i+t^k a^{(k)}_i, &&
\nonumber\\[2pt]
&
\upsilon'^{(k-s)}_i (t',x')=\upsilon^{(k-s)}_i (t,x)+t^s C_k^s a^{(k)}_i, && u'(t',x')=u(t,x), && w'(t',x')=w(t,x),
\end{align}
where $k=0,\dots,2 \ell$, $s \leq k$, $\upsilon^{(0)}_i=x_i$, and $C_k^s$ are the binomial coefficients $C_k^s=\frac{k!}{s! (k-s)!}$.

Comparing (\ref{TR6}) with the infinitesimal form of (\ref{TTR2}) (see  \cite{AG2}), one can express $u$, $w$ in terms of $\rho$ and $\upsilon_i$
\be\label{link}
\rho=e^{-\mathcal{N} \ell u}, \qquad w=\frac{1}{ 2  \mathcal{N} \ell } \rho^{-\frac{1}{ \mathcal{N} \ell}}  \frac{ \partial \upsilon^{(1)}_i }{\partial x_i}.
\ee
Recall that $\upsilon^{(1)}_i$ was earlier identified with the fluid velocity vector field $\upsilon_i$.

Afterwards, one computes the Maurer-Cartan invariants
\be
g^{-1} d g={\rm i} \omega_H  H +{\rm i} \omega_D D +{\rm i} \omega_K K +{\rm i} \omega^{(0)}_i C^{(0)}_i  +{\rm i} \omega^{(1)}_i C^{(1)}_i+\dots+{\rm i} \omega^{(2\ell)}_i C^{(2 \ell)}_i,
\ee
then
attends to an orbit of a liquid particle $x_i \to x_i(t)$, and finally imposes the constraints
\be\label{CON1}
\omega_D=0, \qquad \omega^{(0)}_i=0, \qquad  \omega^{(1)}_i=0, \qquad \dots \qquad \omega^{(2\ell-1)}_i=0,  \qquad \omega^{(2\ell)}_i+\alpha \nabla_i \omega_H=0,
\ee
where $\alpha$ is a real constant and $\nabla_i$ is the invariant derivative (\ref{Na1}). The equations $\omega^{(0)}_i=0$, $\dots$,  $\omega^{(2\ell-1)}_i=0$ allow one to link $\upsilon^{(k)}_i$, with $k=1,\dots,2\ell-1$, to the material derivatives of $x_i$
\be
\upsilon^{(k)}_i=\frac{1}{k!} \mathcal{D}^k x_i,
\ee
with $\mathcal{D}$ specified in (\ref{MD}).
Substituting (\ref{link}) into $\omega_D=0$, one obtains the continuity equation, while the restriction $\omega^{(2\ell)}_i+\alpha \nabla_i \omega_H=0$ reproduces the generalized Euler equation entering Eqs. (\ref{fin}), in which
\be
\nu=\frac{(2\ell)! \alpha }{1+\mathcal{N} \ell}.
\ee
Like in the preceding section, the equation of state exposed in (\ref{fin}) arises automatically.

Note that for an arbitrary value of $\ell$ the explicit form of the Maurer-Cartan invariants is rather complicated.  As an illustration, we expose the $\ell=\frac 32$ case
\bea
&&
\omega_H=e^{-u} dt, \qquad \qquad ~ \omega_D=du-2 w e^{-u} dt, \qquad \qquad ~ \omega_K=dw-w du+ w^2 e^{-u} dt,
\nonumber\\[2pt]
&&
\omega^{(0)}_i=e^{-\frac{3u}{2}}(dx_i-\upsilon^{(1)}_i dt), \qquad \omega^{(1)}_i=e^{-\frac{u}{2}}(d \upsilon^{(1)}_i-2 \upsilon^{(2)}_i dt)- 3 w e^{-\frac{3u}{2}}(dx_i-\upsilon^{(1)}_i dt),
\nonumber\\[2pt]
&&
\omega^{(2)}_i=e^{\frac{u}{2}}(d \upsilon^{(2)}_i-3 \upsilon^{(3)}_i dt)- 2 w e^{-\frac{u}{2}} (d \upsilon^{(1)}_i-2 \upsilon^{(2)}_i dt)+3 w^2 e^{-\frac{3u}{2}}  (dx_i-\upsilon^{(1)}_i dt),
\nonumber\\[2pt]
&&
\omega^{(3)}_i=e^{\frac{3u}{2}} d \upsilon^{(3)}_i- w e^{\frac{u}{2}}(d \upsilon^{(2)}_i-3 \upsilon^{(3)}_i dt)+w^2 e^{-\frac{u}{2}} (d \upsilon^{(1)}_i-2 \upsilon^{(2)}_i dt)
\nonumber\\[2pt]
&&
\qquad \quad -w^3 e^{-\frac{3u}{2}}  (dx_i-\upsilon^{(1)}_i dt).
\eea

To summarize, within the method of nonlinear realizations, both the perfect fluid equations, which hold invariant under the action of the $\ell$-conformal Galilei group, and the equation of state come about naturally. It suffices to consider the group-theoretic element (\ref{GTE}) and then impose the constraints (\ref{CON1}).

\vspace{0.5cm}

\noindent
{\bf 5. Perfect fluid equations with the Lifshitz symmetry}\\

If one omits the special conformal transformation generator $K$ in the Schr\"odinger algebra (\ref{SA}), the commutators $[H,D]$ and $[D,C^{(1)}_i]$ can be modified so as to include an arbitrary constant $z$ known as the dynamical critical exponent (see e.g. \cite{MT})
\bea\label{LA}
&&
[H,D]={\rm i} z  H, \qquad  [H,C^{(1)}_i]={\rm i} C^{(0)}_i, \qquad [D,C^{(0)}_i]=-\frac{{\rm i}}{2} C^{(0)}_i,
\nonumber\\[2pt]
&&
[D,C^{(1)}_i]={\rm i} \left(z-\frac 12 \right) C^{(1)}_i,
\eea
where, as before, $i=1,\dots,\mathcal{N}$.
The resulting algebra is known as the Lifshitz algebra.\footnote{To be more precise, in modern literature by the Lifshitz algebra one usually means (\ref{LA}), in which the generator of Galilei boosts $C^{(1)}_i$ is discarded, while the generators of spatial rotations $M_{ij}={\rm i} \left(x_i \frac{\partial}{\partial x_j}-x_j \frac{\partial}{\partial x_i} \right)$ are reinstated.} It is conventionally represented by the differential operators acting in a nonrelativistic spacetime parameterized by $(t,x_i)$
\be\label{LA1}
H={\rm i} \frac{\partial}{\partial t}, \qquad D={\rm i} z t \frac{\partial}{\partial t} +\frac{{\rm i}}{2} x_i \frac{\partial}{\partial x_i}, \qquad C^{(0)}_i={\rm i} \frac{\partial}{\partial x_i}, \qquad C^{(1)}_i={\rm i} t \frac{\partial}{\partial x_i}.
\ee
In particular, $D$ in (\ref{LA1}) gives rise to the anisotropic scaling of the temporal and spatial coordinates
\be
t'=e^{\lambda z} t, \qquad x'_i=e^{\frac{\lambda}{2}} x_i,
\ee
$\lambda$ being the transformation parameter. The goal of this section is to obtain perfect fluid equations invariant under the action of the Lifshitz group and the corresponding equation of state within the method of nonlinear realizations.

Guided by the analysis in the preceding sections, one directly attends to the group-theoretic element
\be\label{gte3}
g=e^{{\rm i} t H} e^{{\rm i} x_i C^{(0)}_i} e^{{\rm i}  \upsilon_i (t,x) C^{(1)}_i} e^{{\rm i} u(t,x) D},
\ee
and then establishes (finite) transformation laws of the coordinates and fields under the Lifshitz group (each transformation is separated by semicolon)
\begin{align}\label{TRL}
&
t'=t+\beta, && x'_i=x_i, &&
\nonumber\\[2pt]
&
\upsilon'_i (t',x')=\upsilon_i (t,x), && u'(t',x')=u(t,x);
\nonumber\\[6pt]
&
t'=e^{\lambda z}t, && x'_i=e^{\frac{\lambda}{2}}x_i, &&
\nonumber\\[2pt]
&
\upsilon'_i (t',x')=e^{-\lambda\left(z-\frac 12 \right)}\upsilon_i (t,x), && u'(t',x')=u(t,x)+\lambda;
\nonumber\\[6pt]
&
t'=t, && x'_i=x_i+a^{(0)}_i,
\nonumber\\[2pt]
&
\upsilon'_i (t',x')=\upsilon_i (t,x), && u'(t',x')=u(t,x);
\nonumber\\[6pt]
&
t'=t, && x'_i=x_i+t a^{(1)}_i,
\nonumber\\[2pt]
&
\upsilon'_i (t',x')=\upsilon_i (t,x)+a^{(1)}_i, && u'(t',x')=u(t,x),
\end{align}
where $\beta$ and $\lambda$ are (finite) parameters associated with the temporal translation and anisotropic scaling transformation, respectively, while $a^{(0)}_i$, $a^{(1)}_i$ are related to the spatial translations and the Galilei boosts. Then one computes the Maurer-Cartan invariants
\be
g^{-1} d g={\rm i} \omega_H  H +{\rm i} \omega_D D + {\rm i} \omega^{(0)}_i C^{(0)}_i  +{\rm i} \omega^{(1)}_i C^{(1)}_i,
\ee
where
\be
\omega_H=e^{-zu} dt, \qquad \omega_D=du, \qquad
\omega^{(0)}_i=e^{-\frac{u}{2}}(dx_i-\upsilon_i dt), \qquad \omega^{(1)}_i=e^{\left(z-\frac 12 \right)u} d \upsilon_i.
\ee

In order to make contact with fluid mechanics, one introduces the fluid density $\rho(t,x)$ and the velocity vector field $\upsilon_i (t,x)$. By making recourse to (\ref{TRLR}), one finds the transformation law of the density under the dilatation
\be\label{trlr}
\rho(t,x)= e^{\frac{\lambda \mathcal{N}}{2}} \rho' (t',x'),
\ee
while $\rho(t,x)=\rho' (t',x')$ for other transformations from the Lifshitz group. Likewise, considering an orbit of a liquid particle as in Eq. (\ref{vel}) above and attending to (\ref{TRL}), one can identify
$\upsilon_i (t,x)$ in (\ref{gte3}), (\ref{TRL}) with the fluid velocity vector field.

Comparing the transformation laws (\ref{TRL}) and (\ref{trlr}), one can link $u$ to $\rho$
\be\label{RU}
\rho=e^{-\frac{\mathcal{N}}{2} u}.
\ee
The invariant zero-form
\be\label{IZF}
\Omega=-\frac {2}{\mathcal{N}}\rho^{-\frac{2z}{\mathcal{N}}} \frac{\partial \upsilon_i}{\partial x_i},
\ee
and the invariant derivative
\be
\nabla_i=\rho^{-\frac{1}{\mathcal{N}}} \frac{\partial }{\partial x_i}
\ee
can be found as well.

Finally, imposing the constraints built from the invariants
\be\label{conL}
\omega_D+\Omega \omega_H=0, \qquad \omega^{(1)}_i+\alpha \nabla_i \omega_H=0,
\ee
where $\alpha$ is a constant, and taking into account the field redefinition (\ref{RU}), one reduces the first equation in (\ref{conL}) to the continuity equation, while the second condition gives the conventional Euler equation and the equation of state
\be\label{EoS}
\frac{\partial \rho}{\partial t} + \frac{\partial ( \rho \upsilon_i )}{\partial x_i}=0, \qquad \rho \mathcal{D} \upsilon_i =-\frac{\partial p}{\partial x_i}, \qquad
p=\nu \rho^{1+\frac{2(2z-1)}{\mathcal{N}}}.
\ee
Here $\nu$ is a constant, which links to $\alpha$ in (\ref{conL}) via $\nu=\frac{2\alpha z}{\mathcal{N}+2(2z-1)}$.
At $z=1$ one reproduces the result in Sect. 2.

Note that, because each factor contributing to the left hand side of the Euler equation scales in a concrete way under the dilatation transformation, it comes as no surprise that choosing the pressure to be a power function of the density one can fix the exponent so as to ensure the invariance of the entire equation. Similar reasoning worked out in two preceding sections.
To the best of our knowledge, Eqs. (\ref{EoS}) have not yet been discussed in the literature.

A few comments are in order. Taking into account the transformation laws (\ref{TRL}), (\ref{trlr}) and the fact that the material derivative $\mathcal{D}$ introduced in (\ref{MD}) is invariant under each transformation except for the dilatation, for which $\mathcal{D}=e^{\lambda z} \mathcal{D}'$, one can verify that the first two equations in (\ref{EoS}) do hold invariant under the Lifshitz group. One reveals a subtlety in constructing conserved charges, however. Using the relation
\be
\frac{\partial}{\partial t} \left(\int_V dx \rho A \right) =\int_V dx \rho \mathcal{D} A,
\ee
where $A(t,x)$ is an arbitrary function, $\mathcal{D}$ is the material derivative, and $dx=dx_1 \dots dx_{\mathcal{N}}$, which holds true due to the continuity equation and the assumption that $\rho$ vanishes at the boundary of the volume element $V$, one can check that
\be
C^{(0)}_i=\int_V dx \rho \upsilon_i, \quad  C^{(1)}_i=t C^{(0)}_i-\int_V dx \rho x_i, \quad H=\int_V dx \left(\frac 12 \rho \upsilon_i \upsilon_i+V(\rho) \right),
\ee
are conserved over time. Here $V(\rho)$ is the potential, which via the Legendre transform gives the pressure \cite{JNPP}
\be\label{pot}
p(\rho)=\rho V'(\rho)-V(\rho) \quad \Rightarrow \quad  V(\rho)=\frac{\mathcal{N}}{2(2z-1)} p(\rho).
\ee
However, when trying to build a conserved charge associated with the dilatation, which is typically of the form \cite{JNPP}
\be\label{CCD}
D=tH-\frac 12 \int_V dx \rho x_i \upsilon_i,
\ee
one reveals a problem. In view of (\ref{pot}), $D$ in (\ref{CCD}) is conserved for $z=1$ only.

When equations of motion exhibit invariance under a given transformation group, the corresponding action functional may fail to do so. A well known example is the mechanical similarity \cite{LL1}, which is illustrated here by the Lifshitz analog \cite{AG3} of the $1d$ conformal mechanics (\ref{DFF})
\be\label{LifM}
\ddot\rho=\frac{(2 z-1) \gamma^2}{\rho^{4z-1}}.
\ee
Whereas the equation of motion is invariant under the anisotropic conformal transformation
\be
t'=e^{\lambda z} t, \qquad \rho'(t')=e^{\frac{\lambda}{2}}\rho(t),
\ee
the corresponding action functional
\be
S=\frac 12 \int dt \left(\dot\rho^2-\frac{\gamma^2}{\rho^{4z-2}} \right)
\ee
scales as $S'=e^{\lambda(1-z)} S$. Only for $z=1$ the action is invariant and the conserved charge
\be
D=t H-\frac 12 \rho \dot\rho,
\ee
where $H=\frac 12 \left(\dot\rho^2+\frac{\gamma^2}{\rho^2} \right)$ is the energy, can be constructed via Noether's theorem.

It is customary to link conservation of energy and momentum to conservation of the energy-momentum tensor $T^{\mu\nu}$, with $\mu=(0,i)$ and $i=1,\dots,\mathcal{N}$, which for the case at hand reads
\begin{align}
&
T^{00}=\frac 12 \rho \upsilon_i \upsilon_i+V(\rho), && T^{i0}=\rho \upsilon_i \left(\frac 12 \upsilon_i \upsilon_i +V'(\rho) \right), && \partial_\mu T^{\mu 0}=0,
\nonumber\\[2pt]
&
T^{0 i}=\rho \upsilon_i, && T^{ji}=p \delta_{ji}+\rho \upsilon_j \upsilon_i, && \partial_\mu T^{\mu i}=0.
\end{align}
Within the fluid mechanics, the scale invariance is sometimes imposed by representing the dilatation generator in the form $D={\rm i} \xi^\mu \partial_\mu$ and demanding the current $T^{\mu\nu} \xi_\nu$, where $\xi_\nu=\eta_{\nu\mu} \xi^\mu$ and $\eta_{\nu\mu}=\mbox{diag} (+,-,\dots,-)$, to be conserved $\partial_\mu \left( T^{\mu\nu} \xi_\nu \right)=0$. As was explained above, for the case at hand such a current exists for $z=1$ only, whereas the equations of motion (\ref{EoS}) hold invariant for an arbitrary value of $z$.

\vspace{0.5cm}

\noindent
{\bf 6. Perfect fluid equations invariant under relativistic conformal group}\\

We now proceed to studying the case of the relativistic conformal algebra, which is specified by the structure relations
\begin{align}\label{RCA}
&
[D,P_i]=-{\rm i} P_i, && [D,K_i]={\rm i} K_i,
\nonumber\\[2pt]
&
[P_i,K_j]=2 {\rm i} \eta_{ij} D-2 {\rm i} M_{ij}, && [M_{ij},P_l]=-{\rm i} \eta_{il} P_j+{\rm i} \eta_{jl} P_i,
\nonumber\\[2pt]
&
[M_{ij},K_l]=-{\rm i} \eta_{il} K_j+{\rm i} \eta_{jl} K_i, && [M_{ij},M_{kl}]=-{\rm i} \eta_{ik} M_{jl}-{\rm i} \eta_{jl} M_{ik}+{\rm i} \eta_{jk} M_{il}+{\rm i} \eta_{il} M_{jk},
\end{align}
where $i=0, \dots,\mathcal{N}-1$ and $\eta_{ij}=\mbox{diag}(+,-,\dots,-)$ is the Minkowski metric. Above $D$ is identified with the dilatation generator, $P_i$ links to translations in relativistic spacetime, $K_i$ is related to special conformal transformations, while $M_{ij}$ determines the Lie algebra of the Lorentz group. A conventional realization of the generators in Lorentzian spacetime parameterized by coordinates $x^i$ reads
\bea\label{GEN}
&&
D={\rm i} x^i \frac{\partial}{\partial x^i}, \qquad P_i={\rm i} \frac{\partial}{\partial x^i}, \qquad K_i={\rm i} \left( 2 x_i x^j \frac{\partial}{\partial x^j}-x^j x_j \frac{\partial}{\partial x^i}\right),
\nonumber\\[2pt]
&&
M_{ij}={\rm i} \left( x_i \frac{\partial}{\partial x^j}- x_j \frac{\partial}{\partial x^i}\right).
\eea
As usual, the indices are raised and lowered with the use of the Minkowski metric and its inverse. Our objective in this section is to inquire whether the group-theoretic approach is capable of producing relativistic fluid equations with conformal symmetry.

Given the algebra (\ref{RCA}), it seems  natural to start with the group-theoretic element
\be\label{gteF}
g=e^{{\rm i} x^i P_i} e^{{\rm i} \upsilon^i (x) K_i} e^{{\rm i} u(x) D} e^{{\rm i} f^{ij} (x) M_{ij}},
\ee
where $x^i$ are identified with the coordinates parameterizing an $\mathcal{N}$-dimensional Minkowski spacetime and $\upsilon^i (x)$, $u(x)$, $f^{ij} (x)=-f^{ji} (x)$ are fields on it. Proceeding similarly as above, one obtains the infinitesimal conformal transformations acting upon the coordinates and fields (each transformation is separated by semicolon)
\bea\label{TRL7}
&&
x'^i=x^i+\beta^i, \qquad \quad  \upsilon'^i (x')=\upsilon^i (x), \qquad \quad  u'(x')=u(x), \qquad \quad  f'^{ij} (x')=f^{ij} (x);
\nonumber\\[6pt]
&&
x'^i=(1+\lambda)x^i, \qquad \upsilon'^i (x')=(1-\lambda)\upsilon^i (x),  \qquad u'(x')=u(x)+\lambda, \qquad  f'^{ij} (x')=f^{ij} (x);
\nonumber\\[6pt]
&&
x'^i=x^i-\sigma^i (xx)+2 x^i (\sigma x), \qquad \upsilon'^i (x')=(1-2 (\sigma x)) \upsilon^i (x)-2 \sigma^i (x \upsilon)+2 x^i (\sigma \upsilon)+\sigma^i,
\nonumber\\[6pt]
&&
 u'(x')=u(x)+2 (\sigma x),
\qquad e^{{\rm i} f'^{ij} (x') M_{ij}}=e^{-2 {\rm i} x^k \sigma^p M_{kp}} e^{{\rm i} f^{ij} (x) M_{ij}};
\nonumber\\[6pt]
&&
x'^i=x^i-\xi^{ij}x_j, \qquad \upsilon'^i (x')=\upsilon^i (x)-\xi^{ij} \upsilon_j (x), \qquad  u'(x')=u(x),
\nonumber\\[2pt]
&&
e^{{\rm i} f'^{ij} (x') M_{ij}}=e^{\frac{{\rm i}}{2} \xi^{kp} M_{kp}} e^{{\rm i} f^{ij} (x) M_{ij}},
\eea
where $\beta^i$, $\lambda$, $\sigma^i$, $\xi^{ij}=-\xi^{ji}$ are transformation parameters corresponding to translations, dilatations, special conformal transformations, and Lorentz transformations, respectively,  and we abbreviated $(ab)=a^i b_i$. The Maurer-Cartan invariants $g^{-1} dg={\rm i} \omega_D D+{\rm i} \omega^i_P P_i+{\rm i} \omega^i_K K_i+{\rm i} \omega^{ij}_M M_{ij}$
read
\bea\label{MCF}
&&
\omega_D=du-2 (dx \upsilon), \qquad \omega^i_P=e^{-u} dx^j { {\left( \mbox{exp} (-2 f) \right)}_j}^i,
\nonumber\\[2pt]
&&
\omega^i_K=e^u \left( d \upsilon^j-dx^j (\upsilon \upsilon)+2 \upsilon^j (dx \upsilon)\right) { {\left( \mbox{exp} (-2 f) \right)}_j}^i
\nonumber\\[2pt]
&&
\omega^{ij}_M M_{ij}=
\left(d f^{ij} M_{ij}+2 dx^i \upsilon^j e^{-{\rm i} f^{kp} M_{kp}} M_{ij} e^{{\rm i} f^{ls} M_{ls}}\right),
\eea
where ${ {\left( \mbox{exp} (-2 f) \right)}_j}^i$ stands for the conventional matrix exponent.

At this point one reveals a problem. For one thing, the transformation laws of $x^i$ in (\ref{TRL7}) reproduce the generators (\ref{GEN}), which means that the identification of $x^i$ in the group-theoretic element (\ref{gteF}) with the coordinates parameterizing a Lorentzian manifold was correct. For another thing, among the Maurer-Cartan one-forms (\ref{MCF}) one does not find an invariant, which would link $dx^i$ to $\upsilon^i$.
For each nonrelativistic example above, there was such a relation, which helped us to identify $\upsilon_i$ with the fluid velocity vector field.
Furthermore, from (\ref{TRL7}) it follows that $\upsilon^i$ does not transform as a contravariant vector field, i.e. $\upsilon'^i(x') \ne \left( \frac{\partial x'^i}{\partial x^j} \right) \upsilon^j (x)$ with $x'^i=x'^i (x)$ taken from (\ref{TRL7}), and, hence, it cannot be identified with the fluid velocity vector field. Rather, $\omega_D$ suggests that $\upsilon_i$ can be linked to the gradient of $u$. Thus, the construction of relativistic fluid equations invariant under the conformal group in terms of the invariants (\ref{MCF}) alone appears to be problematic.

Yet, discarding $\upsilon_i$ and focusing solely on $x^i$, $u(x)$ and their transformation laws displayed in (\ref{TRL7}), one can build the desired equations. Firstly, one considers the conformal time $s$
\be\label{CT}
ds^2=e^{-2u(x)} \eta_{ij} dx^i dx^j,
\ee
which remains intact under the transformations (\ref{TRL7}). Then one introduces the velocity vector field $V^i (x)$, which is identified with $\frac{d x^i}{ds}$ when restricted upon a liquid particle orbit $x^i=x^i (s)$.
By definition, $V^i(x)$
transforms as a contravariant vector field $V'^i(x')=\left( \frac{\partial x'^i}{\partial x^j} \right) V^j (x)$ if $x'^i=x'^i (x)$ is taken from (\ref{TRL7}). In view of (\ref{CT}), it is constrained to obey
\be\label{CON}
V^i V_i=e^{2u},
\ee
meaning that $V^i$ is a time-like vector. In particular, $V^0$ can be expressed in terms of $V^\alpha$, with $\alpha=1,\dots,\mathcal{N}-1$, and $u$. Note that imposing (\ref{CON}) is consistent with the conformal transformations (\ref{TRL7}).

Taking into account that within the relativistic framework the operator $V^i \partial_i$, where $\partial_i=\frac{\partial}{\partial x^i}$, is an analog of the material derivative (\ref{MD}) and analyzing the way in which  $V^j \partial_j V^i$ is changed under the conformal transformations (\ref{TRL7}), one  finds that introducing two more terms of the type $V^i V^j \partial_j u$ and $e^{2u} \eta^{ij} \partial_j u$ yields the invariant equation\footnote{To be more precise, under the conformal transformations (\ref{TRL7}) the expression on the left hand side of (\ref{REOM}) transforms as a contravariant vector field. Setting it to zero, one obtains the invariant equation.}
\be\label{REOM}
V^j \partial_j V^i-2 V^i V^j \partial_j u+e^{2u} \eta^{ij} \partial_j u=0.
\ee
Note that, in view of (\ref{CON}), a contraction of (\ref{REOM}) with $V_i$ gives zero meaning that only $\mathcal{N}-1$ equations are independent. This correlates with the fact that $V^0$ links to the spatial components $V^\alpha$ and $u$ via (\ref{CON}). In a similar fashion, one can build a relativistic counterpart of the continuity equation
\be\label{RCE}
\partial_i \left( e^{-\mathcal{N} u} V^i\right)=0,
\ee
where $\mathcal{N}$ is the spacetime dimension, which holds invariant under the conformal transformations (\ref{TRL7}).

Finally, introducing the energy-momentum tensor
\be
T^{ij}=e^{-(\mathcal{N}+2)u} V^i V^j-\frac{1}{\mathcal{N}} e^{-\mathcal{N} u} \eta^{ij}, \qquad \partial_i T^{ij}=0,
\ee
which is suggested by Eq. (\ref{REOM}) and focusing on the $T^{00}$-component
\be
T^{00}=e^{-(\mathcal{N}+2)u} V^\alpha V^\alpha+\left(\frac{\mathcal{N}-1}{\mathcal{N}} \right)e^{-\mathcal{N} u},
\ee
where $V^\alpha$, with $\alpha=1,\dots,\mathcal{N}-1$, are the spatial components of the fluid velocity vector field, one concludes that
\be
\varepsilon=\left(\frac{\mathcal{N}-1}{\mathcal{N}} \right)e^{-\mathcal{N} u}
\ee
can be identified with the energy density (measured in the comoving frame). Likewise, analyzing the stress tensor $T^{\alpha\beta}$ in the comoving frame, one determines the pressure
\be
p=\frac{1}{\mathcal{N}} e^{-\mathcal{N} u},
\ee
and establishes the equation of state
\be
p(\varepsilon)=\frac{1}{\mathcal{N}-1} \varepsilon.
\ee
The equations obtained above reproduce the ultrarelativistic limit of the relativistic fluid mechanics \cite{LL}.

\vspace{0.5cm}

\noindent
{\bf 7. Conclusion}\\

To summarize, in this work a possibility to construct perfect fluid equations with conformal symmetry within the method of nonlinear realizations was studied. Four cases were discussed in detail, which included
the Schr\"odinger group, the $\ell$-conformal Galilei group, the Lifshitz group, and the relativistic conformal group. While the method proved rather efficient for the nonrelativistic groups, in particular yielding new results for the Lifshitz case, within the relativistic framework the construction faced certain difficulties and extra arguments needed to be invoked.

Turning to possible further developments, it is interesting to inquire whether thermodynamic description of perfect fluids with conformal symmetry can be accommodated within the group-theoretic framework.

A possibility to combine the method of nonlinear realizations with the approach in \cite{BJLNP} is worth studying as well.

In classical mechanics, having a potential energy, which is a homogeneous function of an argument, results in the mechanical similarity \cite{LL1}, Kepler's third law being the celebrated example. Physical implications of the anisotropic conformal scaling symmetry of the Lifshitz perfect fluid need to be better understood.

A Hamiltonian formulation for the equations of motion in Sect. 4 and 5 is worth exploring as well. Of particular interest here are nontrivial Poisson brackets among the fields and their origin \cite{JNPP}.

\vspace{0.5cm}

\noindent{\bf Acknowledgements}\\

\noindent
The author thanks V.P. Nair for a useful correspondence.

\end{document}